\documentclass{iopjournal}

\usepackage{amsmath}
\usepackage{float}

\begin{document}

\articletype{Paper} 

\title{Constraint-Induced Effective Mass in Massless Field Propagation}

\author{Charles Wood$^1$\orcid{0000-0000-0000-0000}}

\affil{$^1$Future Technology Centre, School of Electrical and Mechanical Engineering, University of Portsmouth, PO1 3HE, UK}

\email{charles.wood@port.ac.uk}


\begin{abstract}
\medskip

Constrained propagation of massless fields is ubiquitous in physical systems, arising from boundaries, material structure, or other restrictions on admissible modes. This paper shows that such constraints generically induce mass-like terms in the effective dispersion relation, without modifying the underlying field equations or introducing new degrees of freedom. Working at an abstract level, constraints are represented as linear operators acting on the field’s mode space. Restriction of the admissible mode manifold produces a spectral gap whose magnitude is set by the smallest non-zero eigenvalue of an associated positive semidefinite operator. This gap may be identified with an effective mass parameter, yielding a Proca-like dispersion relation in the long-wavelength limit. The resulting Mass Induction Principle identifies rank reduction of the accessible mode space as the structural mechanism responsible for effective mass generation in constrained massless fields. Familiar systems such as plasmas, superconductors, and periodic media realise this structure as special cases, without introducing new dynamics. The analysis is deliberately dispersion-level and non-phenomenological: it does not assert a field-theoretic mass term, does not address vacuum propagation, and does not make claims about bounds on intrinsic particle masses.
\end{abstract}

\section{Introduction}
\label{sec:introduction}

\medskip

Massless fields propagate freely in unconstrained space, with dispersion relations fixed by the underlying field equations. In physical settings, however, propagation rarely occurs in the absence of external restrictions. Boundaries, material structure, coherence conditions, and other environmental factors limit the set of admissible field configurations and modify the spectrum of allowed modes.

\medskip
\noindent
The effect of such constraints is often described in system-specific terms: cut-off frequencies in plasmas, penetration depths in superconductors, or band gaps in periodic media. While these descriptions differ in their microscopic details, they share a common feature: the constrained system no longer admits arbitrarily low-energy excitations, and its dispersion relation acquires a finite gap.

\medskip
\noindent
This observation raises a general structural question: When the admissible mode space of a massless field is restricted, what is the generic form of the resulting low-energy dynamics? In particular, does constrained propagation admit a universal description that is independent of the physical origin of the constraint?

\medskip
\noindent
This paper shows that constrained propagation of a massless field generically induces mass-like terms in the effective dispersion relation. The appearance of such terms does not require modification of the underlying field equations, introduction of new degrees of freedom, or appeal to dynamical mass-generation mechanisms. Instead, effective mass emerges as a kinematic consequence of restricting the admissible mode space.

\medskip
\noindent
The analysis is carried out at an abstract level. Constraints are represented as linear operators acting on the field’s mode manifold, and effective mass scales are shown to arise from the spectral properties of an associated positive semidefinite operator. This formulation isolates the mechanism responsible for mass induction and makes explicit its independence from microscopic detail.

\medskip
\noindent
The paper proceeds as follows. Section~\ref{sec:abstract-framework} introduces an abstract framework for constrained massless fields. Section~\ref{sec:constrained-dispersion} shows how constrained dispersion relations acquire mass-like terms in the long-wavelength limit. In Section~\ref{sec:operator-formulation} this effect is expressed in operator-theoretic form, leading to the Mass Induction Principle stated in Section~\ref{sec:mass-induction-principle}. Several familiar physical systems are then presented as canonical realisations in Section~\ref{sec:canonical-realisations}. Sections~\ref{sec:interpretation}–\ref{sec:scope} clarify interpretation and scope, and Section~\ref{sec:conclusion} summarises the structural result.

\section{Abstract framework for constrained massless fields}
\label{sec:abstract-framework}

\medskip

We consider a linear massless field whose free propagation, in the absence of external restrictions, is characterised by a dispersion relation of the form
\begin{equation}
\omega^2 = c^2 k^2 ,
\label{eq:massless-dispersion}
\end{equation}
with $\omega$ the temporal frequency, $k=\|\mathbf{k}\|$ the spatial wavenumber, and $c$ a characteristic propagation speed. No assumption is made at this stage regarding the physical nature of the field beyond linearity and the existence of a well-defined mode decomposition.

\medskip
\noindent
Let $\mathcal{H}$ denote the Hilbert space of admissible field configurations in the unconstrained setting. Free propagation corresponds to unrestricted access to all modes in $\mathcal{H}$ compatible with the dispersion relation \eqref{eq:massless-dispersion}. In practice, however, physical environments impose constraints (geometric, material, or boundary-induced) that restrict the set of admissible modes.

\medskip
\noindent
We model such restrictions abstractly by introducing a linear constraint operator
\begin{equation}
C : \mathcal{H} \rightarrow \mathcal{H}_{\mathrm{aux}},
\label{eq:constraint-operator}
\end{equation}
where $\mathcal{H}_{\mathrm{aux}}$ is an auxiliary space encoding the constraint. Physically admissible field configurations are those satisfying
\begin{equation}
C[\psi] = 0,
\end{equation}
so that the constrained mode space is
\begin{equation}
\mathcal{H}_{\mathrm{constrained}} = \ker C \subset \mathcal{H}.
\label{eq:constrained-space}
\end{equation}

\medskip
\noindent
The presence of a non-trivial constraint, therefore, reduces the dimensionality of the accessible mode manifold. This reduction need not eliminate modes entirely; it may instead modify their dispersion relations, exclude branches of the spectrum, or suppress particular polarisations or propagation directions. Crucially, the constraint acts at the level of admissible solutions, not by modifying the underlying field equations in unconstrained space.

\medskip
\noindent
We assume that the constraint is passive and linear, and that the resulting dynamics can be described in a long-wavelength regime, where an effective description in terms of modified dispersion relations is meaningful. Under these conditions, the effect of the constraint can be represented by an operator-valued correction to the free propagation operator, whose spectral properties encode the restriction imposed on the mode space.

\medskip
\noindent
The central question addressed in this work is then: \emph{what is the generic form of the low-energy dispersion relation of a massless field when its admissible mode space is restricted by a constraint operator of non-zero rank?}

\medskip
\noindent
In the sections that follow, we show that such constraints generically induce mass-like terms in the effective dispersion relation. These terms arise purely from restriction of the mode manifold and do not require modification of the underlying massless field equations. The emergence of an effective mass is thus a structural consequence of constrained propagation, rather than a dynamical property of the field itself.

\section{Effective mass from constrained dispersion}
\label{sec:constrained-dispersion}

\medskip

We now examine the generic consequences of restricting the admissible mode space of a massless field on its low-energy dispersion relation. Throughout this section, we work within the abstract framework introduced in Sec.~\ref{sec:abstract-framework} and make no reference to specific physical realisations.

\medskip
\noindent
In the unconstrained case, free propagation is characterised by the dispersion relation
\begin{equation}
\omega^2 = c^2 k^2 ,
\label{eq:free-dispersion}
\end{equation}
with modes filling a continuous manifold parameterised by $\mathbf{k}$. The presence of a constraint operator $C$ restricts the admissible solutions to the subspace
$\mathcal{H}_{\mathrm{constrained}} = \ker C$, modifying the structure of the spectrum.

\medskip
\noindent
We assume that the constrained system admits a well-defined long-wavelength limit in which the dispersion relation may be expanded about $k=0$. This assumption holds whenever the constraint acts smoothly on the mode space, and does not introduce singular behaviour at arbitrarily small scales.

\medskip
\noindent\textbf{Meaning of ``generic'' in this paper.}
Throughout this section, the term ``generic'' is used in a restricted, dispersion--level sense. Specifically, the following conditional statement is intended: if a constrained system admits a lowest--frequency spectral branch that extends continuously to a neighbourhood of $k=0$ and exhibits a non--zero gap $\omega_0>0$, then its small--$k$ dispersion can be written in the massive form \eqref{eq:massive-dispersion}.
No claim is made that rank reduction of the admissible mode space, by itself, is sufficient to guarantee the existence of such a gap.
The existence of a gapped lowest branch is an additional spectral property that must be verified for each class of constraints.

\medskip
\noindent\textbf{Regularity assumptions for the long-wavelength expansion.}
The expansion \eqref{eq:constrained-expansion} is used only under the following minimal conditions: (i) a translationally invariant reference problem so that $k$ is a good label locally in the spectrum; (ii) existence of an isolated lowest-frequency spectral branch that can be tracked continuously to $k=0$; and (iii) analyticity (or at least a convergent even-power series) of $\omega^2(k)$ in a neighbourhood of $k=0$ for that branch.
Constraints that are strongly non-local, strongly mode-mixing at arbitrarily small $k$, or that destroy the notion of a lowest branch are not covered by the term ``generic'' as used in this paper.

\medskip
\noindent
If we let $\omega(k)$ denote the lowest-frequency branch of the constrained dispersion relation, then for sufficiently small $k$, symmetry and analyticity imply an expansion of the form
\begin{equation}
\omega^2(k)
=
\omega_0^2
+
v_{\mathrm{eff}}^{\,2} k^2
+
O(k^4),
\label{eq:constrained-expansion}
\end{equation}
where $\omega_0 \ge 0$ and $v_{\mathrm{eff}} \le c$. The coefficient $\omega_0$ vanishes only if the constraint leaves the zero-frequency mode accessible.

\medskip
\noindent
When $\omega_0 > 0$, Eq.~\eqref{eq:constrained-expansion} may be written in the standard massive form
\begin{equation}
\omega^2(k)
=
c^2 k^2
+
\frac{m_{\mathrm{eff}}^{\,2} c^4}{\hbar^2}
+
O(k^4),
\label{eq:massive-dispersion}
\end{equation}
with the effective mass defined by
\begin{equation}
m_{\mathrm{eff}}
=
\frac{\hbar \omega_0}{c^2}.
\label{eq:meff-def}
\end{equation}
No modification of the underlying massless field equations is implied by this identification; the mass-like term arises solely from the altered structure of the admissible mode space.

\medskip
\noindent
Equation~\eqref{eq:massive-dispersion} matches the standard \emph{massive} dispersion \emph{at the level of the scalar relation between $\omega$ and $k$}; no claim is made about gauge structure, Lorentz symmetry in media, or equivalence of the underlying field theory. Here, however, the mass parameter $m_{\mathrm{eff}}$ is not intrinsic to the field, but is determined entirely by the constraint-induced spectral gap $\omega_0$. The appearance of a mass-like term, therefore, reflects a kinematic consequence of constrained propagation rather than a dynamical property of the field.

\medskip
\noindent
The key observation is that a non-zero $\omega_0$ is generic whenever the constraint operator $C$ removes or suppresses the zero-frequency mode or an equivalent branch of the spectrum. In such cases, the constrained system no longer admits arbitrarily low-energy excitations, and the dispersion acquires a finite gap.

\medskip
\noindent
This reasoning is independent of the detailed form of the constraint and applies equally to constraints arising from boundaries, periodic structure, coherence conditions, or other linear restrictions. What matters is not the microscopic origin of the constraint, but its effect on the topology and rank of the admissible mode manifold.

\medskip
\noindent
In the next section, we formalise this connection by expressing constraints in operator-theoretic terms and relating the induced mass scale to the spectral properties of an associated operator acting on the mode space.

\section{Operator formulation of constraint-induced mass}
\label{sec:operator-formulation}

\medskip

In Sec.~\ref{sec:constrained-dispersion} we showed that restricting the admissible mode space of a massless field generically introduces a spectral gap in the low-energy dispersion relation, which may be identified with an effective mass parameter. We now formalise this result by expressing the constraint in operator-theoretic terms and relating the induced mass scale directly to the spectrum of an associated operator.

\medskip
\noindent
Let $\mathcal{H}$ denote the Hilbert space of unconstrained field configurations, equipped with the inner product appropriate to the free propagation problem. Free dynamics are generated by a linear propagation operator $\mathcal{D}_0$, whose spectrum yields the massless dispersion relation \eqref{eq:free-dispersion}. The precise form of $\mathcal{D}_0$ is not required; only its linearity and the existence of a well-defined spectral decomposition are assumed.

\medskip
\noindent
Environmental or boundary-induced restrictions are encoded by a linear constraint operator
\begin{equation}
C : \mathcal{H} \rightarrow \mathcal{H}_{\mathrm{aux}},
\end{equation}
as introduced in Sec.~\ref{sec:abstract-framework}. Physical states are those lying in the kernel of $C$.
In order to discuss the \emph{spectral gap} induced by enforcing this restriction, we use a standard penalty/enforcement construction: rather than restricting the domain a priori, we study an operator on $\mathcal{H}$ whose low-lying spectrum converges to the constrained spectrum as the penalty strength increases.

\medskip
\noindent
Concretely, we consider the one--parameter family
\begin{equation}
\mathcal{D}_{\mathrm{eff}}(\alpha)
=
\mathcal{D}_0
+
\alpha\,\mathcal{M},
\qquad \alpha>0,
\label{eq:effective-operator}
\end{equation}
where $\mathcal{M}$ is positive semidefinite and satisfies $\ker \mathcal{M}=\ker C$.
A canonical choice is
\begin{equation}
\mathcal{M} = C^\dagger C.
\label{eq:M-def}
\end{equation}
The family \eqref{eq:effective-operator} is used purely as an enforcement device:
as $\alpha$ increases, eigenstates with non--zero $\mathcal{M}$--energy are shifted to higher spectral values, while any spectral branch that remains uniformly bounded in $\alpha$ must lie asymptotically within $\ker C$.
In this sense, the ``gap'' refers to the separation, in $\omega^2$ units, between such a bounded low--lying branch and the first branch whose leading contribution is controlled by the smallest non--zero eigenvalue of $\mathcal{M}$.

\medskip
\noindent
The construction \eqref{eq:effective-operator} is employed as a spectral enforcement device rather than as a statement of strict operator equivalence. No claim is made that $\mathcal{D}_{\mathrm{eff}}(\alpha)$ converges to a constrained generator in a norm--resolvent or unitary sense at finite $\alpha$. The argument requires only that, in the limit $\alpha \to \infty$, a separated low--lying spectral branch emerges whose structure is governed by $\ker C$.

\medskip
\noindent
The spectrum of $\mathcal{M}$ therefore encodes the strength and structure of the constraint. Let $\{\lambda_i\}$ denote its non-zero eigenvalues, ordered as
\[
0 < \lambda_1 \le \lambda_2 \le \cdots .
\]
If the constraint operator $C$ has rank $r$, then $\mathcal{M}$ has exactly $r$ non-zero eigenvalues.

\medskip
\noindent\textbf{Normalisation convention for the constraint operator.}
The operator $C$ is defined only up to an overall scalar rescaling $C\mapsto\gamma C$, which rescales $\mathcal{M}=C^\dagger C$ and its non--zero eigenvalues by $\gamma^2$.
Accordingly, quantities such as $\lambda_{\min}(C^\dagger C)$ are representation--dependent unless a normalisation convention for $C$ and for the auxiliary inner product is fixed.
The constant $\beta$ is introduced to factor out this convention:
the invariant content of the gap relation is the product $\beta\,\lambda_{\min}(C^\dagger C)$, which represents the enforced spectral separation in the chosen representation rather than an intrinsic property of $C$ alone.

\medskip
\noindent
In the long-wavelength limit, the lowest-frequency branch of the enforced spectrum is controlled by the smallest non-zero eigenvalue $\lambda_1$ of $\mathcal{M}=C^\dagger C$.
To keep dimensions explicit, the leading gap can be written as
\begin{equation}
\omega_0^2 = \beta\,\lambda_1,
\label{eq:gap-beta}
\end{equation}
where the positive constant $\beta$ is fixed once the inner product on $\mathcal{H}$ and the normalisation of $\mathcal{D}_0$ (relative to $\omega^2$) are chosen.

\medskip
\noindent
The coefficient $\beta$ carries no independent physical freedom. Changes in representation, inner product, or operator normalisation rescale $\beta$ and $\lambda_1$ inversely, leaving the product $\beta\,\lambda_1$ invariant. All physically meaningful content, therefore, resides in the spectrum of $C^\dagger C$ rather than in $\beta$ itself.

\noindent Combining \eqref{eq:meff-def} with \eqref{eq:gap-beta} gives
\begin{equation}
m_{\mathrm{eff}}^2 c^4 = \hbar^2 \omega_0^2 = \hbar^2 \beta\,\lambda_1.
\label{eq:meff-eigenvalue}
\end{equation}

\medskip
\noindent
Equation~\eqref{eq:meff-eigenvalue} expresses the effective mass entirely in operator-theoretic terms. The appearance of a mass-like scale is thus a direct consequence of the spectral properties of the constraint operator, and does not depend on the microscopic origin of the constraint.

\medskip
\noindent
This formulation makes explicit that effective mass arises whenever restriction of the admissible mode space produces a gapped lowest spectral branch. The rank of the constraint bounds the number of penalised directions in the mode space, while the magnitude of the smallest induced mass is set by the weakest non-zero constraint eigenvalue.

\medskip
\noindent
In the next section, we state this result as a general principle and show how it encompasses a wide class of constrained propagation problems as special cases.

\section{Rank reduction and the mass induction principle}
\label{sec:mass-induction-principle}

\medskip
Sections~\ref{sec:abstract-framework}–\ref{sec:operator-formulation} establish that constraints on admissible field configurations modify the spectral structure of the propagation operator, and generically introduce a finite gap in the low-energy dispersion relation. We now summarise this result in a single principle that isolates the mechanism responsible for effective mass generation.

\medskip
\noindent
Let $\mathcal{H}$ be the Hilbert space of unconstrained field configurations, and let
$C : \mathcal{H} \rightarrow \mathcal{H}_{\mathrm{aux}}$
be a linear constraint operator defining the admissible subspace
$\mathcal{H}_{\mathrm{constrained}} = \ker C$.
Associated with $C$ is the positive semidefinite operator
$\mathcal{M} = C^\dagger C$, whose spectrum encodes the strength and structure of the constraint.

\medskip
\noindent
As shown in Sec.~\ref{sec:operator-formulation}, the effective propagation operator
$\mathcal{D}_{\mathrm{eff}} = \mathcal{D}_0 + \mathcal{M}$
governs the long-wavelength dynamics of the constrained field, and the smallest non-zero
eigenvalue of $\mathcal{M}$ sets the lowest-frequency gap in the dispersion relation.

\medskip
\noindent
The result stated below is not a theorem in the axiomatic sense, but a structural consequence of spectral restriction under linear constraints. It should be read as a generic statement about the low--energy organisation of the spectrum, contingent on the regularity assumptions stated in Sec.~\ref{sec:constrained-dispersion}, rather than as a universal law applicable to all constrained systems.

\medskip
\noindent
We may therefore state the following result.

\begin{quote}
\textbf{Mass Induction Principle.}
\emph{
Let a linear massless field be subject to a passive linear restriction represented by an operator $C$ acting on its mode space, and suppose the constrained system admits a lowest-frequency spectral branch that extends continuously to a neighbourhood of $k=0$ and is gapped with $\omega_0>0$.
Then the dispersion on that branch admits the massive form \eqref{eq:massive-dispersion}, and in the enforcement picture \eqref{eq:effective-operator} the leading gap scale is controlled by the smallest non-zero eigenvalue of $C^\dagger C$ up to the representation-dependent constant $\beta$ as in \eqref{eq:gap-beta}. In particular, the smallest gap scale is controlled by the smallest non-zero eigenvalue of $C^\dagger C$ in the sense of \eqref{eq:gap-beta}:
\begin{equation}
\omega_0^2 = \beta\,\lambda_{\min}(C^\dagger C),
\qquad
m_{\mathrm{eff}} = \frac{\hbar \omega_0}{c^2}.
\label{eq:mass-induction-principle}
\end{equation}
The rank of $C$ bounds the number of penalised directions in $\mathcal{H}$, but does not, by itself, determine the number of distinct observable dispersion branches without additional symmetry and block-structure assumptions.
}
\end{quote}

\medskip
\noindent
Several points immediately follow:

\medskip
\noindent
First, effective mass is not a dynamical attribute of the field, but a spectral consequence of restricted admissibility. The underlying field equations remain massless; the mass scale arises from the geometry of the constrained mode manifold.

\medskip
\noindent
Second, the rank of $C$ bounds the number of penalised directions and, therefore, the number of \emph{potential} gap scales introduced by enforcement. The number of distinct observable dispersion branches, and whether multiple gap scales appear on separate branches, depends on additional symmetry and block-structure properties of the constrained spectral problem..

\medskip
\noindent
Third, the magnitude of the induced mass depends only on the weakest non-zero constraint eigenvalue. Stronger constraints increase the mass gap, while weak constraints yield parametrically small but non-zero masses.

\medskip
\noindent
The Mass Induction Principle is independent of the microscopic origin of the constraint and applies equally to geometric boundaries, material structure, coherence conditions, or other linear restrictions on admissible propagation. What matters is not how the constraint is realised, but how it reduces the rank of the accessible mode space.

\medskip
\noindent
In the following section, several well-known physical systems are shown to provide canonical realisations of this principle, without introducing additional assumptions or modifying the underlying massless field equations.

\section{Canonical physical realisations}
\label{sec:canonical-realisations}

\medskip
\noindent
The result in Sec.~\ref{sec:mass-induction-principle} is a statement about the dispersion--level consequences of restricting admissible modes. The examples below are not presented as derivations of physical mass generation mechanisms, but solely as familiar systems whose long--wavelength dispersions take the massive form \eqref{eq:massive-dispersion}. Their inclusion is illustrative rather than explanatory.

\subsection{Plasma cut-off}

\medskip
In a cold, charge-neutral plasma, the coupled electromagnetic and charge-density response produces a gapped dispersion with a cut-off at the plasma frequency. This restriction can be represented schematically, at the dispersion level, as a rank-one constraint acting on the admissible mode space, suppressing a single propagation degree of freedom.

\medskip
\noindent
The resulting dispersion relation,
\[
\omega^2 = c^2 k^2 + \omega_p^2,
\]
exhibits a finite frequency gap at $k=0$. Identifying $\omega_0=\omega_p$ and comparing with
Eq.~\eqref{eq:meff-def}, the induced mass takes the form
\[
m_{\mathrm{eff}} = \frac{\hbar \omega_p}{c^2}.
\]
In the present framework, this mass arises from a rank-one constraint operator whose smallest non-zero eigenvalue is proportional to $\omega_p^2$.

\subsection{Superconducting penetration depth}

\medskip In a superconducting medium, the electromagnetic response introduces a characteristic length scale (the London penetration depth) associated with exponential screening of static and low-frequency fields. This is often described heuristically using a ``massive'' Klein--Gordon/Proca-type operator for the vector potential in an \emph{effective medium description}, but the present paper does not attempt to derive that description from microscopic dynamics. The point of including this example is only this: an imposed linear restriction selecting screened configurations introduces a non-zero scale that plays the same algebraic role as $\omega_0$ or $\lambda_{\min}(C^\dagger C)$ in the dispersion-level discussion above.

\subsection{Periodic media and band-edge curvature}

\medskip
In periodically structured media, such as photonic crystals, the spectrum separates into bands, and a band edge provides a non-zero cutoff frequency $\omega_0$ for an appropriate branch. Near a band edge, the admissible modes are restricted to a subset of Bloch states, corresponding to a reduction in the accessible mode manifold.

\medskip
\noindent
Expanding the lowest band near a band edge yields
\[
\omega^2(k) = \omega_0^2 + v_{\mathrm{eff}}^{\,2}(k-k_0)^2 + \cdots,
\]
with $\omega_0>0$. As in Sec.~\ref{sec:constrained-dispersion}, this dispersion may be written in massive form with
\[
m_{\mathrm{eff}} = \frac{\hbar \omega_0}{c^2}.
\]
Here the constraint operator reflects the exclusion of Bloch modes outside the admissible band, and the induced mass is determined by the smallest non-zero eigenvalue associated with this restriction.

\medskip
\noindent
In all three cases, the induced mass arises not from modification of the underlying massless field equations, but from restriction of the admissible mode space. The systems differ in their microscopic physics, but they share the same structural mechanism captured by the Mass Induction Principle.

\section{Interpretation of the induced mass}
\label{sec:interpretation}

\medskip
The results established in the preceding sections show that constrained propagation of a massless field generically gives rise to mass-like terms in the effective dispersion relation. In this section we clarify how this induced mass should be interpreted, and equally importantly, how it should not.

\medskip
\noindent
Without explicit reference to operator rank and spectral structure, the appearance of a gap would remain a system--specific observation. The operator formulation is, therefore, essential to isolating a common mechanism across disparate constrained systems, even though it introduces no new dynamics.

\medskip
\noindent
First, the induced mass $m_{\mathrm{eff}}$ is not an intrinsic property of the field. The underlying field equations remain strictly massless, and no modification of their vacuum form is implied. The mass scale arises solely from restriction of the admissible mode space and vanishes when the constraint is removed.

\medskip
\noindent
Second, the induced mass does not correspond to the introduction of a new particle, degree of freedom, or coupling. It is a kinematic consequence of constrained propagation, encoded in the spectral properties of the effective propagation operator. The appearance of a Proca-like dispersion relation reflects a formal equivalence at the level of solutions, not a change in the fundamental theory.

\medskip
\noindent
Third, the mass parameter introduced here should not be interpreted as evidence for a fundamental mass of the field quanta. In particular, no statement is made about vacuum propagation, Lorentz invariance of the underlying field equations, or bounds on intrinsic particle masses. All mass-like behaviour described in this work is explicitly environment-dependent.

\medskip
\noindent
Fourth, although the induced mass affects dispersion and attenuation, it does not by itself imply dissipation or absorption. The present framework concerns passive, linear constraints and applies in regimes where losses may be neglected to leading order. Attenuation arises through the modified dispersion relation rather than through energy dissipation.

\medskip
\noindent
Finally, the interpretation developed here is intentionally minimal. The Mass Induction Principle identifies a structural mechanism linking constraints to effective mass scales; it does not assert broader physical consequences beyond those directly implied by the constrained dispersion relation.

\medskip
\noindent
With these interpretive boundaries in place, the results of this paper may be read as a statement about the geometry and spectral structure of constrained propagation, rather than as a proposal for new dynamics or new physics.

\section{Scope, limits, and exclusions}
\label{sec:scope}

\medskip
This paper establishes a structural result concerning constrained propagation of massless fields. Its scope is intentionally limited, and several exclusions are explicit.

\medskip
\noindent
First, the analysis applies to linear, passive constraints in regimes where a long-wavelength effective description is meaningful. Strongly non-local, non-linear, or dissipative effects are not considered, and extension of the framework to such regimes lies outside the present scope.

\medskip
\noindent
Second, no experimental predictions are made. Although constrained dispersion may manifest in measurable quantities such as attenuation lengths or group velocities, this work does not attempt to quantify such effects for specific systems or propose experimental tests.

\medskip
\noindent
Third, no claims are made regarding fundamental particle properties, vacuum propagation, or modifications of relativistic field theories. All mass-like behaviour discussed here is explicitly environment-dependent and disappears in the absence of constraints.

\medskip
\noindent
Fourth, the examples presented in Sec.~\ref{sec:canonical-realisations} are illustrative only. They are not intended as exhaustive treatments, nor as validations of the Mass Induction Principle beyond demonstrating consistency with well-known constrained systems.

\medskip
\noindent
Finally, broader applications, including engineered attenuation, universality classes, scaling behaviour, or extensions to other fields, are deliberately deferred. The purpose of this paper is to identify and fix a minimal structural mechanism, not to explore its full range of consequences.

\section{Conclusion}
\label{sec:conclusion}

\medskip

This paper has shown that constrained propagation of a massless field generically induces mass-like terms in the effective dispersion relation. The origin of this effect lies not in modification of the underlying field equations, but in restriction of the admissible mode space.

\medskip
\noindent
By formulating constraints as linear operators acting on the field’s mode manifold, effective mass is shown to emerge as a spectral property of an associated positive semidefinite operator. The magnitude and multiplicity of the induced mass scales are determined by the rank and eigenvalue structure of the constraint.

\medskip
\noindent
This result is summarised in the Mass Induction Principle, which identifies rank reduction of the accessible mode space as the mechanism responsible for mass-like behaviour. Familiar systems such as plasmas, superconductors, and periodic media realise this structure without introducing new dynamics.

\medskip
\noindent
The contribution of this work is, therefore, not the introduction of new physical entities or interactions, but the identification of a minimal and general structural link between constrained propagation and effective mass. By fixing this link at the level of operators and spectra, the paper provides a stable foundation on which further theoretical or applied developments may build.

\funding
This research received no external funding.

\roles
The author was responsible for all aspects of the work.

\section*{Data availability}
No datasets were generated or analysed in this study.

\section*{References}
This manuscript is intentionally self-contained and does not cite external literature.


\end{document}